\documentclass[aps,pre,preprint,groupedaddress,showpacs]{revtex4}   
\usepackage{graphicx}
\usepackage{bm}
\usepackage{setspace}                         
\usepackage{amsmath}
\usepackage{longtable}
\newcommand{\kmax}{$k_{max}$}
\newcommand{\jmax}{$j_{max}$}
\newcommand{\cth}{$c_{th}$}
\newcommand{\mth}{$m_{th}$}
\newcommand{\rl}{recurrence length}
\newcommand{\lmch}{$l^{m}_{ch}$}

\newcommand{\pacz}{BAPA}
\newcommand{\calntwrk}{CALNTWRK}

\begin{document}

\title{Earthquake Correlations and Networks--- A Comparative Study}
\author{Krishna Mohan, T. R.}
\email{kmohan@cmmacs.ernet.in}
\homepage{http://www.cmmacs.ernet.in/~kmohan}
\author{Revathi, P. G.}
\email{revathi.pg@gmail.com}
\affiliation{CSIR Centre for Mathematical Modelling and Computer Simulation (C-MMACS)\\Bangalore 560017, India}
\date{\today}

\begin{abstract}
We quantify the correlation between earthquakes and use the same to distinguish between relevant causally connected earthquakes. Our correlation metric is a variation on the one introduced by Baiesi and Paczuski (2004). A network of earthquakes is constructed, which is time ordered and with links between the more correlated ones. Recurrences to earthquakes are identified employing correlation thresholds to demarcate the most meaningful ones in each cluster. Data pertaining to three different seismic regions, viz. California, Japan and Himalayas, are comparatively analyzed using such a network model. The distribution of recurrence lengths and recurrence times are two of the key features analyzed to draw conclusions about the universal aspects of such a network model. We find that the unimodal feature of recurrence length  distribution, which helps to associate typical rupture lengths with different magnitude earthquakes, is robust across the different seismic regions. The out-degree of the networks shows a hub structure rooted on the large magnitude earthquakes. In-degree distribution is seen to be dependent on the density of events in the neighborhood. Power laws, with two regimes having different exponents, are obtained with recurrence time distribution. Thisis in agreement with the Omori law for aftershocks and extends it to spatial recurrences. The crossover to the second power law regime can be taken to be signalling the end of aftershock regime in an objective fashion.

\end{abstract}

\maketitle
\section{Introduction}
Many complex procesess in nature have the intrinsic tendency of self-organizing themselves into a critical state. For example, earthquakes, forest fires, avalanches, biological evolution etc. have this feature in common. These processess are characterized by long range correlations in space and time and typically also exhibit power law distributions in many of their variables~\cite{Newman2007}. However, the dynamics of the process that self-organizes  the system to the critical state  is still not understood clearly ~\cite{DaPcz2005}.  In recent times, many studies  have sought to understand seismicity from the viewpoint of complex networks \cite{AbeSuz2004, AbeSuz2006,   BaPcz2004, Pcz2005, Jdetal2006, Grassetal2007}. The spatiotemporal properties of seismicity of a region are analyzed from the patterns exhibited by the network constructed from the earthquake catalog of the region. Such an approach which focusses on the spatial and temporal links between   nodes represented by the  events in the catalog, without  considereing the causes of such linkages, is particularly useful in seismicity where the underlying dynamics is still obscure. We adopt this method in this study and compare different networks contsructed from earthquake catalogs from three different seismic regions of the globe, viz. California, Japan and Himalayas.

The clustering of earthquakes in space and time suggests that events that follow in time are, to a certain extent, causally related to the earlier ones. However, restricting the causality connection to a single predecessor or to the somewhat arbitrary mainshock-aftershock scenario may not be enough. Rather, the causality connection can be extended to a cluster of events that are strongly correlated based on data analysis. With this in mind, Baiesi and Paczuski \cite{BaPcz2004} introduced a metric to quantify correlations between earthquakes. We use a similar metric here but with a change which will be clarified below. The metric is obtained by combining two of the most robust statistical laws that  characterize earthquake data--- the Gutenberg-Richter law \cite{GR1941} and the fractal distribution of earthquake epicenters (see, for example, \cite{Turcot1997}). The former law states that the number, $N$, of earthquakes of magnitude $\ge m$ vary as \begin{equation} \label{Gut} N(m) \sim 10^{-bm} \end{equation} where $b$ is a constant $\approx 1$, but does vary a little with the region and the catalog (see, for example, \cite{BaYiOz}). The latter is based on fractal analysis and is of recent origin, but has been shown to be quite robust through the analysis of various datasets from various regions of the globe (see, for example, \cite{Hirata, Turcot1997}): \begin{equation} \label{fract} N'(l) \propto l^{d_f} \end{equation} where $N'$ is the number of pairs of points separated by a distance $l$ and $d_f$ is the fractal dimension.

If we combine the above two laws, we may state that the average number of events that can occur in the region within a distance $l_{ij}$ separating two events $i$ and $j$ is \begin{equation} \label{nij}  n_{ij} = Kl^{d_{f}}_{ij}10^{-bm_{i}}\Delta m_{i} \end{equation} where $K$ is a constant of proportionality that may be related to the seismic activity of the region \cite{Ba2006}, and $\Delta m_{i}$ is the accuracy in measurement of the magnitude $m_{i}$. Note that the events of the catalog are assumed to be time ordered with $j > i$ and $t_j > t_i$. We may then define a correlation relation between earthquakes as  \begin{equation} \label{eqcor} c_{ij} = \frac{1}{n_{ij}} \end{equation} and note that the correlation between two events is a maximum when $n_{ij}$ is at a minimum.

We construct the network now by making the linkages between correlated events in the time ordered list of earthquakes. For this purpose, correlation thresholds, \cth, are defined and only those event pairs with $c_{ij} \ge$ \cth\ are linked together in the network. The basic premise is that all subsequent events in the catalog are recurrences to the selected event, but a pruning is being done based on \cth\ to ensure that only the most correlated are selected.

Our metric for measuring correlation is similar to the one used by Baiesi and Paczuski \cite{BaPcz2004} (hereafter referred to as \pacz), but, we will be  focussing on spatial recurrences rather than aftershocks. Thus, we have chosen to omit the time factor $t_{ij} \ \ (= t_j - t_i)$ from the expression occurring in \pacz.  The metric used here ensures that spatially closeby events are more correlated and, also, a higher magnitude event contributes a larger value to correlation than a lower magnitude one at that same location. In particular, our metric does not, unlike in \pacz, leave out events that are separated by large time intervals, as long as they satisfy the correlation relation, Eq.~\ref{eqcor} (see Krishna Mohan and Revathi \cite{Rev-Calif} for a more detailed comparison).

We studied, in a recent work~\cite{Rev-Calif} (hereafter referred to as \calntwrk), the spatial recurrences of earthquakes in the California region by following the above procedure. In this  paper we adopt the same procedure to  construct networks of correlated events from an earthquake catalog of Japan  and from a catalog for the Himalayan belt. We  analyze the  seismicity of these regions in terms of the topology of these networks and compare the same with the findings from \calntwrk\ to extract the robust features of such correlated networks of seismic events.

\section{The Region and Catalogs} The regions chosen for the study here are both regions of high seismic potential.  Japan is located in a region of considerable seismic risk. Japan lies on the cusp of the Pacific-Philippine-Eurasian triple plate junction, where the complex interactions of three tectonic plates is unpredictable and loaded with potential activity.  Seismicity is dominated by the subduction of the Pacific plate  under the Okhotsk Plate  to the north and, in  Southern Japan, the subduction of the Philippine Sea plate under the Amurian Plate and the Okinawa Plate~\cite{air}. The complex interaction of these plates has produced a long history of damaging earthquakes. The recurrence interval of earthquakes along most crustal faults in Japan is typically quite long, while the recurrence intervals of events along subduction zones is usually much shorter~\cite{herp}. The region covered in this analysis lies between $(126.433^{\circ}\mathrm{E}$--$148.0^{\circ}\mathrm{E})$ longitudes and $(25.730^{\circ}\mathrm{N}$--$47.831^{\circ}\mathrm{N})$ latitudes.

The Himalayas are one among the most seismically active regions of the world. The seismicity of the Himalayas is contributed mainly by the north-south convergence of the Indian and  Eurasian plate, the east-west  convergence of the Indo-Burmese mountain and the underthrusting of the Indian plate below the Eurasian plate~\cite{lakshmi}. Several studies of this region indicate that shallow focus earthquakes dominate this region and the fault plane solutions indicate  the dominance of thrust faulting and strike slip in this region~\cite{lakshmi}. Though the whole Himalayan belt may be considered as one seismic belt, we have carried out the analysis based on a division into three zones--- Western Himalayas (WH) (between $(70^{\circ}\mathrm{E}$--$78^{\circ}\mathrm{E})$ longitudes and $(30^{\circ}\mathrm{N}$--$38^{\circ}\mathrm{N})$ latitudes), Central Himalayas (CH) (between $(78^{\circ}\mathrm{E}$--$98^{\circ}\mathrm{E})$ longitudes and $(28^{\circ}\mathrm{N}$--$38^{\circ}\mathrm{N})$ latitudes) and North Eastern Himalayas (NEH) (between $(88^{\circ}\mathrm{E}$--$98^{\circ}\mathrm{E})$ longitudes and $(20^{\circ}\mathrm{N}$--$28^{\circ}\mathrm{N})$ latitudes). Due to paucity of data, CH is not analyzed here.

\begin{figure}
\caption{The $\log$-$\log$ plot of Eq.~\ref{Gut} for the different regions studied here; the panels have been labelled after the corresponding seismic regions, `Jap' for Japanese data (Fig.~\ref{GR}(a)), `WH' for Western Himalayas  (Fig.~\ref{GR}(b)) and `NEH' for North Eastern Himalayas (Fig.~\ref{GR}(c)). The minimum magnitude, $m_{min}$, used in these plots were varied and the  minimum value of $m_{min}$ that gives a reasonable linear behavior ascertained. $b$ value of the corresponding plot and the data corresponding to that plot, which is what is displayed here,  were subsequently employed in the analysis presented here.}
\includegraphics[]{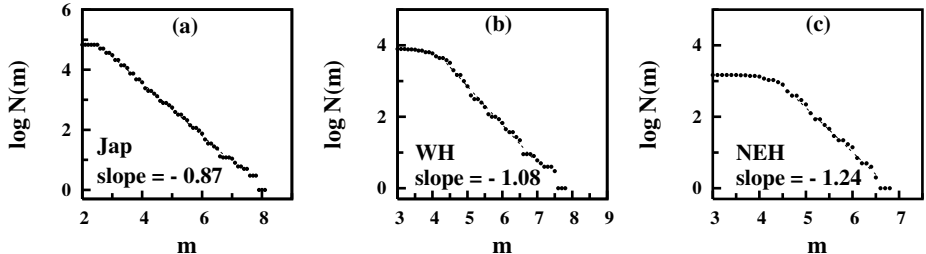}
\label{GR}
\end{figure}
The data sources are, (1)~Japan  University Network Earthquake Catalog ~\cite{jap_catalog} for the Japanese data (January 1, 1993 to December 31, 1998) and, (2)~Advanced National Seismic System (ANSS) catalog (January 1, 1973 to December 31, 2007) for the Indian Himalayas. The data was analyzed against Eq.~\ref{Gut} to confirm adherence to the GR law, by varying the minimum magnitudes, $m_{min}$,  considered in the data sets. The minimum of $m_{min}$ which gives a reasonable power law behavior was subsequently chosen (Fig.~\ref{GR}) and only data with $m > m_{min}$ considered for the analysis presented here. The  $m_{min}$ chosen was  3.0 for the Japanese data and  4.5 for the Indian Himalayas. The Japanese data uses the JMA scale to express magnitude while the ANSS data is expressed in moment magnitude. While the ANSS catalog is for a much longer period, however, the minimum magnitude for the catalog is higher at 4.5 and, hence, the number of events is much lesser, and thus the data set much poorer.

 \section{Methodology}

We have chosen to retain the values for $K$ and $\Delta m_{i}$ as was used in \calntwrk\   for all the regions studied here. The values of $b$ were obtained from a linear fit to the the $\log$-$\log$ plots of Eq.~\ref{Gut} (Fig.~\ref{GR}) and were  $0.87$ for Japan, $1.24$ for WH and $1.24$ for NEH.  $d_f$ was set equal to   1.6 for Japan~\cite{Hirata} while, for WH and NEH, we estimated $d_f$ ourselves as 1.6 and 1.4 respectively. 

\subsection{Distribution of Correlation values}
\begin{figure}[h]
\includegraphics[]{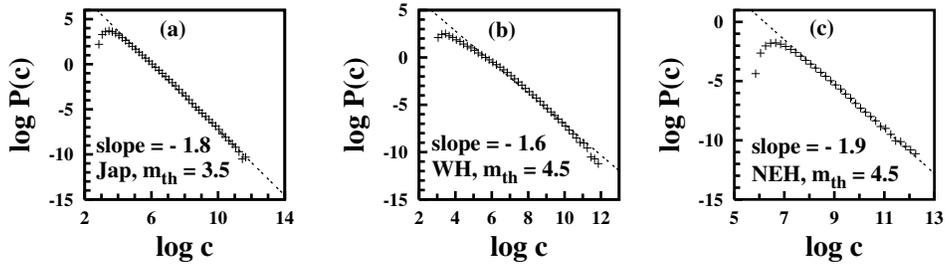}
\caption{The distribution of correlation values, evaluated using Eq.~\ref{eqcor}, for all  pairs of events of the corresponding earthquake catalogs. The panels have been labelled after the catalogs; Fig.~\ref{correl}(a) is for Japan, Fig.~\ref{correl}(b) is for WH and Fig.~\ref{correl}(c) is for NEH.  It is seen that the correlation values follow a power law across many decades. The range of the power law is seen to be more with a lower \mth\ value (Fig.~\ref{correl}a). }
\label{correl}
\end{figure}
The distribution of correlation values between all  pairs of events in each catalog is similar. Excepting at the upper and lower limits, the distribution is a power law over many decades of the log scale. The range of the power law is more if the \mth\  value of the catalog is smaller, as can be seen from Fig.~\ref{correl}.  The distribution for Japan which has a \mth\ $= 3.0$ shows a range of about 7 decades for the power law regime while the two Himalayan regions with  \mth \ $= 4.5$  have ranges $< 5$ decades. \calntwrk\ had \mth\ $= 2.5$ and a range of about $8$ decades. 

The  exponent of the power law for all the regions are in the range $-1.6$ to $-1.9$; \calntwrk\  had an exponent of $-1.7$.  However, whether  the exponent depends on the region can only be established if we compare the values obtained from catalogs with similar degrees of completeness,  homogeneity and accuracy. For example, we noticed that, in general, for the same region and same catalog, if we increase \mth, the exponent of the power law decreases. Nevertheless, for different regions, the exponent does not agree for the same \mth\ values.

In the analysis below, we have tested the sensitivity of our results as \cth\ and \mth\ values are varied. For this purpose, based on Fig.~\ref{correl}, it was decided to explore a range of \cth\ values from  $10^{6} $ to $10^{9}$ for the Japanese data. Also, the different \mth\ values used in the analysis of the Japanese data are 3.0, 3.5, 4.0 and 4.5. Likewise, we employed a range of \cth\ values from $10^{8}$ to $10^{11}$ for the WH data and $10^{7}$ to $10^{10}$ for the NEH data. For both of these data sets, the \mth\ values tried were 4.5, 5.0 and 5.5. 

\section{Results}

\subsection{Degree distributions}

\begin{figure}[h]
\includegraphics[]{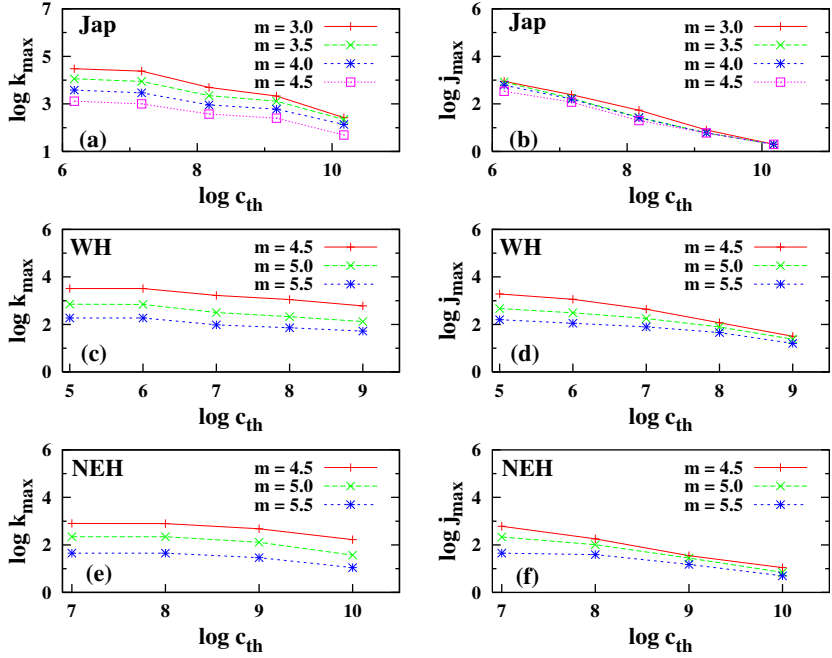}
\caption{The maximum out-degree (\kmax) (left panels) and maximum in-degree (\jmax) (right panels) values are plotted against increasing \cth\ values for the different seismic regions considered. The seismic regions pertaining to the panels have been  labelled therein with Fig.~\ref{kmax}(a) and Fig.~\ref{kmax}(b) for Japan, Fig.~\ref{kmax}(c) and Fig.~\ref{kmax}(d) for WH and Fig.~\ref{kmax}(e) and Fig.~\ref{kmax}(f) for NEH. In general, \kmax\ $>$ \jmax\ for the corresponding \mth\ values, for all \cth\ values and, in most cases, by at least an order of magnitude. This indicates that a hub structure is followed as far as out-degree is concerned and the in-degree distribution is density dependent. The difference between \kmax\ and \jmax\ is much less, though, in the case of NEH (Fig.~\ref{kmax}(f)). The \jmax\ values converge to a common value as \cth\ is increased.  The \kmax\ values fall off as \cth\ is increased, but the graphs for the different \mth\ values are parallel and not convergent, indicating a scale invariance with respect to magnitude.}
\label{kmax}
\end{figure}

The maximum out-degree (\kmax) and maximum in-degree (\jmax) distributions for the Japanese  seismic network shows (Fig.~\ref{kmax}(a) and Fig.~\ref{kmax}(b) respectively) almost similar features as \calntwrk. In particular, a hub structure is present as far as out-degree is concerned while it is absent with the in-degree distribution; in-degree distribution is more density dependent. The former can be deduced from the fact that \kmax\ values are consistently much higher than the \jmax\ values. \kmax\  is, for the Japanese data, at least an order of magnitude higher than the \jmax\  for the corresponding \mth\ values. Large earthquakes will tend to have more events associated with them through higher correlation values since Eq.~\ref{eqcor} is positively correlated with magnitude. On the other hand, lower magnitude events require small $l_{ij}$'s to obtain larger correlation values, i.e. the density of points become important for such events. In the case of both the Himalyan seismic regions (Fig.~\ref{kmax}(c) and Fig.~\ref{kmax}(d) respectively for WH and, Fig.~\ref{kmax}(e) and Fig.~\ref{kmax}(f) respectively for NEH), we observe that the difference in corresponding \kmax\ and \jmax\ values are slightly less. This needs to be investigated with better data sets to try to understand whether the distribution of different magnitudes across the region is in any significant way different from the other regions. 

As \cth\ is increased, \kmax\ values fall-off in approximately a power law fashion. This was observed with \calntwrk\  and we see the same behavior with these data sets too. However, the fall-off in the Japanese case is slightly less uniform. Whereas with the \calntwrk, we observed a slower initial fall-off, followed by a faster fall-off for the higher \cth\ values, we find this pattern repeated twice for the Japanese data. The parallelism between the different graphs for different \mth\ values during the fall-off, which confirms that a scale invariance with respect to magnitude is maintained as \cth\ is increased, is mostly present with the Japanese data and, excepting only the graphs for \mth\ $= 3.0$ and  \mth\ $= 3.5$  which converge to the same value for \cth\ $= 10^{10}$. 

\begin{figure}[h]
\includegraphics[]{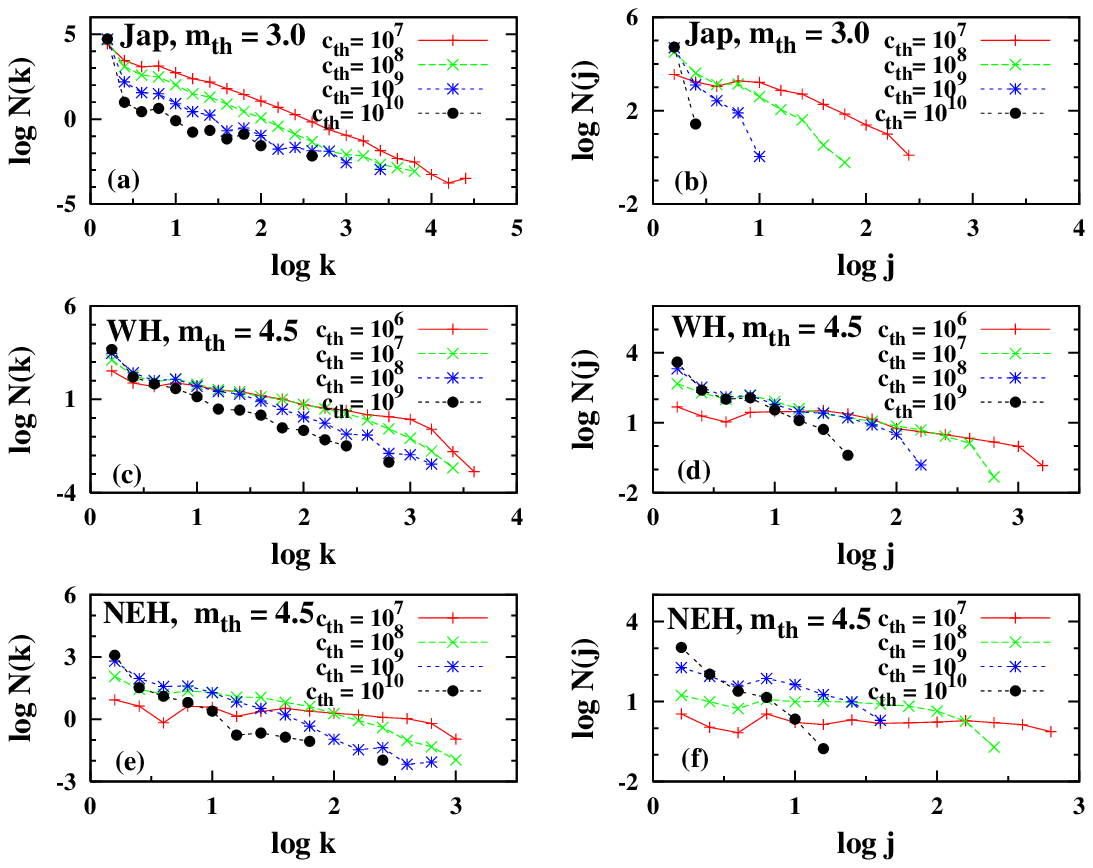}
\caption{The general out-degree ($k$) (left panels) and in-degree ($j$) (right panels) distributions are plotted for the different seismic regions with the panel labels indicating the region analyzed. Fig.~\ref{degree}(a) and Fig.~\ref{degree}(b) are for Japan, Fig.~\ref{degree}(c) and Fig.~\ref{degree}(d) are for WH and Fig.~\ref{degree}(e) and Fig.~\ref{degree}(f) are for NEH. It is seen that the $k$ distribution is, on the average, a power law with departures from it for low and high $k$ values. The only other departure from this behavior is with NEH (Fig.~\ref{degree}(e)) which shows a significant plateau in the graph for \cth\ $= 10^7$, indicating a uniform density of points for a wide range of $k$ values. The $j$ distribution shows a plateau for low \cth, which converts to a power law fall-off as \cth\ is increased. The plateau portion is quite reduced in the case of Japanese data (Fig.~\ref{degree}(b)).}
\label{degree}
\end{figure}
We had observed, in the case of \calntwrk, a convergence of \jmax\ values as \cth\  is increased resulting in a common value for all \mth\ values at higher \cth\ values. This is in accordance with the observation that the in-degree is controlled by the density of points and, as \cth\ is increased, only the higher magnitude events and closeby events are left in the clusters and, given the sparse network,  all clusters have more or less similar densities. We observe the same behavior here with all data sets. The only difference is that, in the case of Japanese data, the \jmax\ values, for different \mth\ values, are not very different even for low \cth\  values (Fig.~\ref{kmax}(b)). This uniformity in the density of \jmax\ values across different \mth\ values needs to be explored further because it suggests a uniform distribution of magnitudes with respect to \jmax. 

As far as the general out-degree distribution is concerned (Fig.~\ref{degree}(a) for the Japanese data, Fig.~\ref{degree}(c) for WH and Fig.~\ref{degree}(e) for NEH), all the regions considered have similar behaviors with an approximate power law fall-off with increasing $k$ values. For intermediate range of $k$ values, the exponent is close to $-2.0$ for the \calntwrk\ and Japanese data. This is also in agreement with the exponent value quoted by \pacz. On an average (for \cth\ $> 10^7$), same exponent value is obtained for the Himalayan region too. For \cth\ $ = 10^7$, a departure is observed from this value in the case of WH and NEH and, in particular, NEH shows a significant plateau region (see below). The graphs depart from the general trend for very small $k$ values as well as very large $k$ values. This is understandable since very small $k$ values indicate isolated events and very large $k$ values indicate very large magnitude events which are also sparse given the GR law. There is, however, some new features present in the $k$ distribution for the Himalayas. This is prominent in the case of NEH where we see that, for the smallest \cth\ value, a significant plateau is observed in the distribution. We can identify a similar tendency in the case of WH even if it is not so prominent. Since smaller \cth\ values are dominated by smaller magnitudes and larger distances, we see in these cases a tendency towards a uniform density  with respect to the spatial distribution of such events. A more complete and homogeneous data set needs to be studied with respect to these areas to clarify this behavior further.

The general in-degree distribution (Fig.~\ref{degree}(b) for the Japanese data, Fig.~\ref{degree}(d) for WH and Fig.~\ref{degree}(f) for NEH) is similar across all regions of California, Japan and the Himalayas. They all show a significant plateau region for the lowest \cth\ value suggesting a uniformity with respect to the distribution of intermediate $j$ values across all the clusters, a prominent density dependent feature. When the network gets sparser with higher \cth\ values, the number of available $j$ values decrease, and the distribution has a steeper fall off in an approximate power law fashion.

\subsection{Recurrent time distributions}
An important feature of earthquake data is that aftershocks decay according to the Omori law ~\cite{omori1894, utsu1995} \[ n(t) \sim \frac{K}{(c+t)^p} \] where $c$ and $K$ are constants in time but depend on magnitude $m$ and, $p \approx 1$;  $n$ is the number of aftershocks per unit time. Earthquakes of all magnitudes have aftershocks  which decay according to the Omori law. It is widely recorded in literature that even intermediate magnitude events can have aftershocks that persist upto years~\cite{BaPcz2004}. We investigate this by studying the distribution of {\em recurrence times} (also referred to as `waiting times' in the literature \cite{Baketal2002, AlCorral2003}) in the data.

\begin{figure}
\includegraphics[]{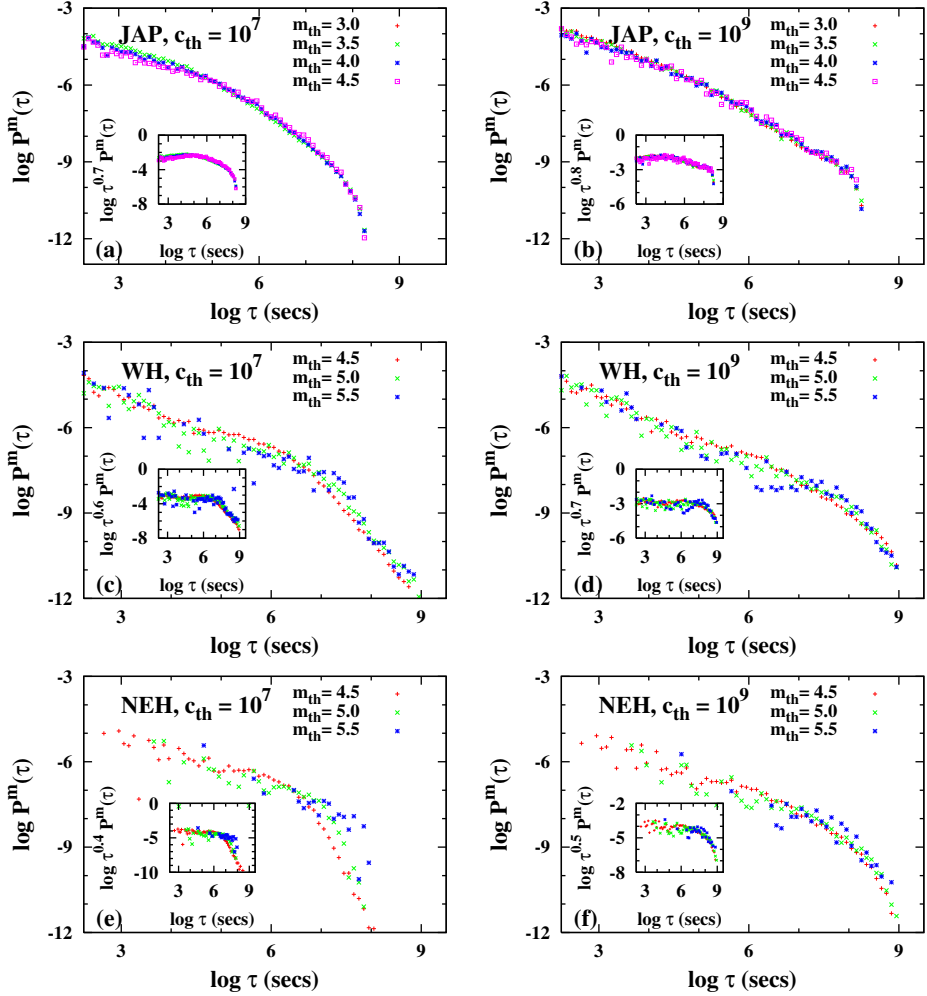}
\caption{Recurrence time ($\tau$) distribution is plotted for two different \cth\ values for each of the three different seismic regions (one in each row). The regions have been labelled in the panels with Fig.~\ref{rtimes}(a) and Fig.~\ref{rtimes}(b) for Japan, Fig.~\ref{rtimes}(c) and Fig.~\ref{rtimes}(d) for WH and Fig.~\ref{rtimes}(e) and Fig.~\ref{rtimes}(f) for NEH.  In general, a power law distribution is observed, but with two regimes having different exponents. The insets to the plots shows a rescaled plot which brings out the linearity of the $\log$-$\log$ plot better, with the two different linear regimes more easily seen. The larger \cth\ value (right panels) brings out the power law feature better, even if the data paucity increases, in particular for the Himalayan regions.}
\label{rtimes}
\end{figure}

We define here the recurrence time, $\tau$, as the time between successive events in a cluster surrounding an event.  $\tau$ ranges from seconds to tens of years in all the clusters. We have left out $\tau$ $< 180$ seconds because of the likely error margins. In the case of \calntwrk, we had observed that, in a $\log$-$\log$ plot of the distribution of $\tau$, a power law holds for most of the range of recurrence times and, in particular, for about 5--6 decades. Furthermore, the graphs for the different \mth\ values overlap indicating that the same law holds well for all $m_{th}$  values. In the case of Japanese data (Fig.~\ref{rtimes}(a) and Fig.~\ref{rtimes}(b)), we see that such a situation holds very well (for about 4 decades in this case) when \cth\ is higher at $10^9$ (Fig.~\ref{rtimes}(b)). For a lower \cth\ value of $10^6$ (Fig.~\ref{rtimes}(a)), we see that the $\log$-$\log$ plot departs from the straight line behavior of a power law and is more parabolic in shape. 

In the insets provided with the plots, we have shown rescaled graphs which bring out the linear aspect better. From the inset of Fig.~\ref{rtimes}(b), we see that the appearance of about 4 decades of linearity in the main graph is not fully correct. The plot has, rather, two linear regimes, the first one from about $\log \tau = 3$ to $5$ and the second from  $\log \tau = 5$ to $8$. A similar procedure of rescaling, in Fig.~\ref{rtimes}(a), shows two linear regimes from  $\log \tau = 3$ to $5$ and, from  $\log \tau = 5$ to $8$. 

The Himalayan data, due to its sparsity, shows a lot of scatter and does not give visually appealing graphs (Fig.~\ref{rtimes}(c) and Fig.~\ref{rtimes}(d) for WH and Fig.~\ref{rtimes}(e) and Fig.~\ref{rtimes}(f) for NEH). Nevertheless, we can again identify a similar trend of two linear regimes in the $\log$-$\log$ plots. We revisited the data for \calntwrk\  and concluded that a two linear regimes analysis does hold in that case too even though the first linear regime is quite prominent in that case and lasts for almost 5 decades.

Let us recollect that the spatial recurrences analyzed here includes more events than aftershocks which are restrained to be within a `reasonable' (quite often chosen arbitrarily)  time window after the main event. As such, what we observe is that the power law behavior seen with respect to the Omori law can be extended to recurrences as well. However, the exponent of the power law changes after a certain point in time which can be used to identify the limit of the time window within which the aftershocks lie. This, then, provides an objective method of choosing the aftershocks window as opposed to the arbitrary methods in practice now.

\subsection{Recurrence length distributions}
By recurrence length, we mean the distance from an event to each of the other events in the same cluster. In other words, we are measuring the scatter in the way the subsequent events in a cluster are distributed around the event in question.  While there is no established law like the Omori law for recurrence lengths, studies have indicated that a power law seems to hold for recurrence lengths as well \cite{Grassetal2007}. To investigate this, we use  the great circle distance $r_{ij}$ for computing the distance between two event locations. This is calculated using the  Haversine formula \cite{haversine}: if $(\phi_s, \lambda_s)$ and  $(\phi_f, \lambda_f)$ are the (latitude, longitude) values for the two event locations and, $\Delta \phi = \phi_f - \phi_s$ and  $\Delta \lambda = \lambda_f - \lambda_s$, then \[ r_{ij} = R \Delta \sigma\] where $\Delta \sigma$, the  spherical distance, or central angle, between the two event locations is \[ \Delta \sigma = 2 \arcsin \left( \sqrt{\sin^2 \left(\frac{\Delta \phi}{2}\right) + \cos(\phi_s) \cos(\phi_f) \sin^2 \left(\frac{\Delta \lambda}{2}\right)} \right). \] $R$, the radius of the earth,  is taken as $6367~ km$.

\begin{figure}[h]
\includegraphics[]{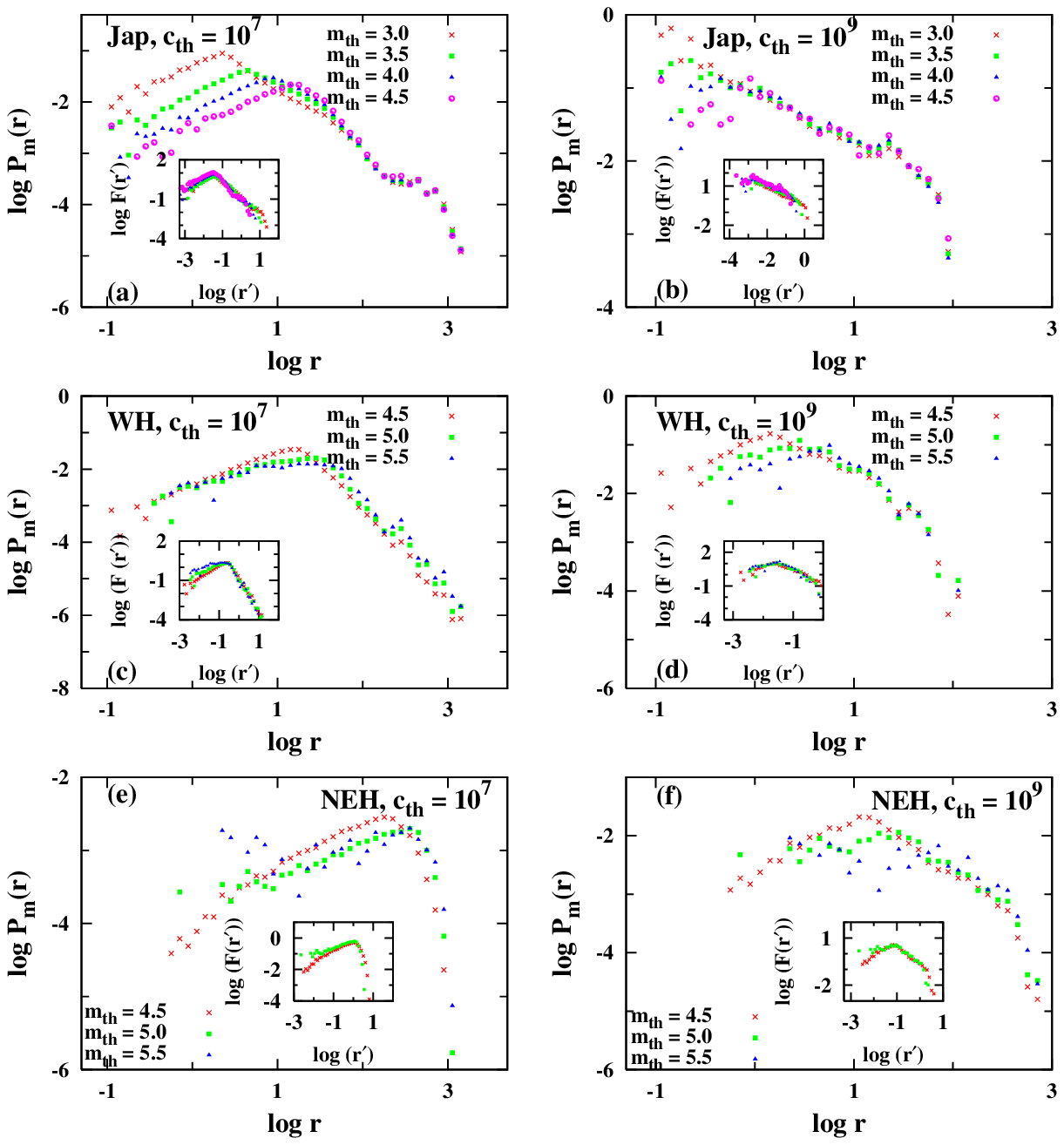}
\caption{Recurrence length ($r$) distribution is plotted for two different \cth\ values (left and right panels) for the three different seismic regions considered here (in the the three rows) which are indicated in the panels. Fig.~\ref{rlplot}(a) and Fig.~\ref{rlplot}(b) are for Japan, Fig.~\ref{rlplot}(c) and Fig.~\ref{rlplot}(d) are for WH and Fig.~\ref{rlplot}(e) and Fig.~\ref{rlplot}(f) are for NEH. The distribution is unimodal in all cases with the peak shfiting to higher $r$ values as \mth\ is increased. This suggests that the peak value (\lmch) may be associated with a characteristic rupture length for the different  magnitudes. The graphs for the different \mth\ values, for each \cth\ value, are superposed after the peak and well separated and, in general, parallel before the peak. This may be understood as due to the fact that the data for the lower \mth\ value contains the data for the higher \mth\ value. As \cth\ value increases, the part before the peak gets shortened (compare right panels with the corresponding left panels), eventually disappearing for sufficiently higher \cth\ values (not shown here). The insets show  rescaled graphs, where   $r' = r/10^{\alpha m_{th}}$  and $F(r') = 10^{\alpha m_{th}} P_{m}(r)$; $\alpha$ is a function of the data set and is computed from a linear fit to the plot of \lmch\ values against \mth. We see that all the graphs for the different \mth\ values are collapsed onto a single curve with a single peak.}
\label{rlplot}
\end{figure}
The general appearance and behavior of the graphs for Japanese as well as Himalayan data (Fig.~\ref{rlplot}) are similar to the \calntwrk. The probability distribution of recurrence lengths is unimodal in nature and peaks at a \rl\ ($r^{m}_{ch}$) which is larger for increasing values of $m_{th}$, for a fixed $c_{th}$. The sparsity of Himalayan data gives rise to much scatter and does not give clean graphs (Fig.~\ref{rlplot}(c) and Fig.~\ref{rlplot}(d) for WH and Fig.~\ref{rlplot}(e) and Fig.~\ref{rlplot}(f) for NEH). Nevertheless, the above behavior can be clearly seen there too. Keep in mind that the distribution with lower \mth\ value contains the data points appearing in the distribution with higher \mth\ values. Also, not all the data points appearing with the lower \mth\ value is present with the higher \mth\ value. Thus, as \mth\ increases, the data points corresponding to the lower $m$ values get left out of the distribution. The shifting of \lmch\ to the right, as \mth\ is increased, therefore points to the presence of a characteristic \lmch\ associated with particular $m$ values and this clearly is larger for larger $m$ values. Since the rupture is larger with larger earthquakes, we may think of \lmch\ as representative of a typical rupture length associated with events of that magnitude (see also \cite{Grassetal2007}).
 
While the graphs for different $m_{th}$ values are separated before the rise to the peak, they are collapsed onto a single curve after their peaks. We had observed, in our study of \calntwrk, that the part of the graphs, for the different \mth\ values, are parallel, while being separated, before the peak. This trend is easily enough observed in the case of Japanese data, but more difficult to confirm for the Himalayan data because of scatter induced by data paucityepresence of the power law, before the peak,  with a common exponent for the different \mth\ values, suggest that the location errors for these small \rl\ values are not random in nature and, if any, are systematic in nature.  This is contrary to what has been claimed in literature \cite{Grassetal2007}, as already suggested by us in the earlier work \cite{Rev-Calif}. 

As \cth\ is increased, the part of the graphs before the peak is shortened (compare the right panels with the corresponding left panels in Fig.~\ref{rlplot}), eventually disappearing for sufficiently high \cth\ values (not shown here).  Note that we are plotting only pairs of events with $r_{ij} >$ 100 m. As such, the peaks disappear for larger values of $c_{th}$ since the peak values are now $<$ 100 m. The behaviour seen with \lmch\ is completely in accordance with Eq.~\ref{nij}. As \cth\ increases, we are selecting event pairs of larger magnitudes separated by smaller distances. 

The part of the graphs, for different \mth\ values, after the peaks, are collapsed onto a single curve till data paucity forces scatter for the higher \rl\ values. The former may be expected from the fact, already stated, that the data points for the lower \mth\ value contain the data points for the higher \mth\  values. In the case of Japanese data (Fig.~\ref{rlplot}(a) and Fig.~\ref{rlplot}(b)), we see a small plateau appearing after the fall-off just before the very high \rl\ values. In the Himalayan case (Fig.~\ref{rlplot}(c)--Fig.~\ref{rlplot}(f)),  the fall-off is seen to be much steeper compared to the other cases, indicating that larger recurrence lengths are fewer. 

The  plots for the  different values of \mth\   collapse on to a single curve when the $x$ axis is rescaled as $r' = r/10^{\alpha m_{th}}$  and the $y$ axis by  $10^{\alpha m_{th}} P_{m}(r)$ with $\alpha$ a function of the data set and, computed from a linear fit to the plot of \lmch\ values against \mth. $\alpha$ was seen to be the same regardless of the \cth\ value in the case of \calntwrk. We observe the same feature in the case of Japanese data as well. However, in the case of Himalayan data, different $\alpha$ values were obtained for the different \cth\ values. This was not surprising because, given the sparse data set and, with just three \mth\ values to play with, it was difficult to spot the \lmch\ peaks correctly. We have, however, still used a single  $\alpha$ value for any particular Himalayan region, with this value being an averaged value over the different $\alpha$ values obtained for the  different \cth\ values. 

\section{Conclusions}
In a continuation of our earlier study, we have compared the network features of seismic data from three different areas, viz. California, Japan and Himalayas. The network features correspond to the networks constructed according to our algorithm, based on the modification of an algorithm given by \pacz. Both these algorithms are based on a metric for correlations between earthquakes and our algorithm differs from \pacz\  in omitting the time factor from their definition to enable the recurrences to be identified, as opposed to the identification of aftershocks by \pacz.

Our comparisons show that there are some quite robust features present in networks constructed from seismic data from around the globe. These include a hub structure in the out-degree distribution accompanied by a density dependent in-degree distribution, unimodal \rl\ distribution with a peak value (\lmch) that increases with \mth\ indicating that characteristic rupture lengths for different magnitudes  may be present and, a recurrence time distribution which has two power law regimes with different exponents. We have suggested, with respect to the latter, that the recurrence time at which the first power law regime changes to the second one be used, as an objective criterion, for the upper limit of the time window used for selecting aftershocks. 

It is clear, from this comparative study, that earthquake networks provide us with a useful analysis tool to study the seismicity patterns of the globe. Further work in this area should lead us to a better understanding of the event distribution and, eventually, the dynamics behind it.

\bibliographystyle{abbrv}
\bibliography{recurrence_compare}

\end{document}